\documentclass[journal, 10pt]{IEEEtran}
\usepackage[nocompress]{cite}
\usepackage[dvips]{graphicx}
\graphicspath{{./eps/}}
\DeclareGraphicsExtensions{.eps}
\usepackage{amsmath}
\usepackage[ruled]{algorithm2e}
\usepackage{array}
\usepackage[caption=false,font=footnotesize,labelfont=sf,textfont=sf]{subfig}
\usepackage{stfloats}
\usepackage{url}
\usepackage{balance}
\usepackage{hyphenat}
\usepackage{multirow}
\begin{document}
%
%

\title
	{
	Selecting Microarchitecture Configuration of Processors for Internet of Things
	}
%
%
\author
  {
  Prasanna~Kansakar,%
  ~\IEEEmembership{Student~Member,~IEEE}
  and~Arslan~Munir,%
  ~\IEEEmembership{Senior Member,~IEEE}

  \thanks{The authors are with the Department of Computer Science, Kansas State University, Manhattan, KS\protect\\
e-mail: \{pkansakar@ksu.edu,~amunir@ksu.edu\}}
	}
\maketitle
\sloppy
\nohyphens{	
%
%
\begin{abstract}
The Internet of Things (IoT) makes use of ubiquitous internet connectivity to form a network of everyday physical objects for purposes of automation, remote data sensing and centralized management/control. IoT objects need to be embedded with processing capabilities to fulfill these services. The design of processing units for IoT objects is constrained by various stringent requirements, such as performance, power, thermal dissipation etc. In order to meet these diverse requirements, a multitude of processor design parameters need to be tuned accordingly. In this paper, we propose a temporally efficient design space exploration methodology which determines power and performance optimized microarchitecture configurations. We also discuss the possible combinations of these microarchitecture configurations to form an effective two-tiered heterogeneous processor for IoT applications. We evaluate our design space exploration methodology using a cycle-accurate simulator (ESESC) and a standard set of PARSEC and SPLASH2 benchmarks. The results show that our methodology determines microarchitecture configurations which are within 2.23\%--3.69\% of the configurations obtained from fully exhaustive exploration while only exploring 3\%--5\% of the design space. Our methodology achieves on average 24.16$\times$ speedup in design space exploration as compared to fully exhaustive exploration in finding power and performance optimized microarchitecture configurations for processors.
\end{abstract}
%
%
%
%
\section{Introduction and Motivation}
\label{introduction_and_motivation}
%
\IEEEPARstart{T}{he} internet has grown rapidly in both enterprise and consumer markets. This has given rise to the Internet of Things (IoT) wherein everyday physical objects are interconnected through a communication network for purposes of automation, remote data sensing and centralized management/control. The IoT creates an intelligent, invisible network fabric that can be sensed, controlled and programmed which  allows objects in IoT ecosystem to communicate, directly or indirectly, with each other or the Internet \cite{tiEvolutionIoT2013}. The ``things", in the scope of IoT, are IoT enabled objects containing sensing and actuating elements along with embedded hardware and software components which facilitate data aggregation, network connectivity and security. Each IoT enabled object is designed to perform an application specific task using data gathered by itself or using information made available to it through other objects in the network. There has been widespread deployment of IoT objects in recent years in various applications like healthcare, industry, transportation etc. It is estimated that 6.4 billion connected end-devices are in use in the year 2016 \cite{gartner6BillionEst2015}, with the number expected to rise to 26 billion by the year 2020 \cite{tiEvolutionIoT2013}.

The massive deployment of IoT objects results in generation of large volumes of data. Data communication, processing, real-time analysis and security of such large volumes of data are important issues that need to be resolved for efficient growth of the IoT ecosystem in the years to come. In the current IoT model, IoT end-devices are designed to be as simple and as cost effective as possible. Thus, they are designed with limited processing capabilities, just enough to securely connect and offload data to the cloud. Almost all complex data management functionalities such as data filtering and analysis are delegated to cloud datacenters, the core units of the IoT model. With the growth in data volume in the IoT ecosystem, there rises several significant challenges which renders this model infeasible. We list here three such challenges.

\begin{itemize}
\item \emph{Network Overload} - Core network bandwidth is a vital resource in the IoT ecosystem which must be used efficiently. With ever increasing number of IoT objects, relaying data over the core network to the cloud, the network is severely overloaded. Network overloads introduce latency in critical data processing operations which impact most IoT applications such as healthcare and transportation that require real time data processing.

\item \emph{Data security} - Data communication in the IoT ecosystem mostly occurs over the public network infrastructure. In order to ensure secure data communication, several complex security protocols must be applied to the data. The volume of data requiring security increases as the number of IoT objects deployed in the IoT ecosystem increases. Applying complex security protocols to large volumes of data requires extensive computing operations which cannot be matched by the energy budget of IoT objects.

\item \emph{Upgradability} - As the IoT landscape continues to evolve, it becomes necessary to upgrade IoT deployments in frequent periods. IoT objects must be designed to support hassle free addition of new features via remote access. In an ideal IoT model, IoT objects must be able to upgrade to new, more complex features without deployment of new IoT objects and without any direct human involvement. With limited processing ability, addition of new features to existing IoT objects may be challenging or even infeasible.
\end{itemize}

The challenges posed by the current IoT model can be overcome by adding processing capabilities inside or local to IoT objects \cite{intrinsycHeteroComp2016}. With the added processing units, data management operations such as filtering and analysis can be carried out within the local network. IoT objects can thus, communicate summaries of information, obtained from filtering the aggregated data, to the cloud. This contributes significantly to freeing up the core network bandwidth. The reduction in data volume also reduces the energy expenditure on data security as less data requires lesser number of computing operations to secure. Having more processing ability also makes IoT deployments more flexible to upgrades as newer features can be added without significantly burdening the system.

Processing units interfaced with IoT objects require an optimal balance between power and performance \cite{intelDevSolIoT2014}. Since many IoT objects are battery powered, it is desirable that these objects operate for their entire lifetime with the battery they are deployed with (e.g. medical sensors implanted into a patient's body via invasive surgical process). Although great progress has been made in battery technology, batteries are still not able to keep pace with the demands of modern electronics \cite{mgEmbSysPwrSwHw2011}. So, power optimization must be considered in parallel with performance optimization.

\begin{figure}[!h]
\centering
\includegraphics[width = 3.5in, bb = 7 7 1299 676]{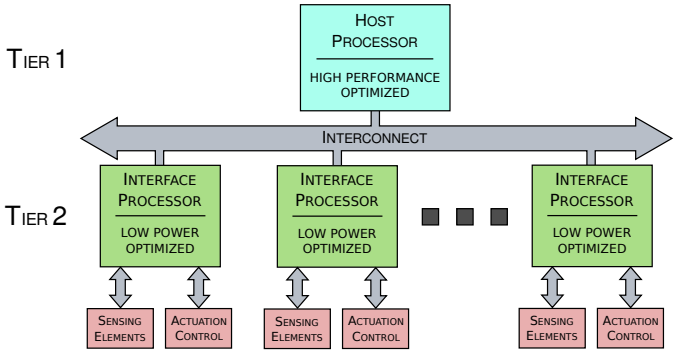}
\caption{Two-tiered heterogeneous processor architecture model for IoT}
\label{figure_heterogeneous_processor}
\end{figure}

For incorporating higher levels of power optimized performance in IoT deployments, a two-tiered heterogeneous processor architecture is suitable \cite{intrinsycHeteroComp2016}\cite{armFlexIoTNodes2015}. This two-tiered architecture, shown in Figure \ref{figure_heterogeneous_processor}, consists of a host processor, optimized for high performance, interfaced with a number of interface processors, optimized for low power operation. The interface processors collect data from data-sensing elements and control actuating elements. These processors are always operated in active mode because their low power operation does not severely impact battery life. Higher end function, such as filtering and analysis of data, and, implementation of complex security protocols are performed by the host processor. Since these operations are infrequent, the power hungry host processor is mostly operated in sleep state and only activated intermittently for limited durations.

Designing efficient embedded processors with power-optimized performance, for use in IoT objects, is a tedious process. Preventing high performance processors from violating the power budget requirements dictated by the market is an enormous design challenge \cite{mgPowerESL2010}. The opportunities for optimizing a processor design for power are the greatest at the architecture level \cite{mgPowerESL2010}. Thus, power and performance optimizations should be performed while defining the microarchitecture configuration of processors. The microarchitecture configuration consists of several processor design parameters each of which has to be tuned based on the impact it has on the overall power and performance of the processor. Selecting a microarchitecture configuration involves rigorous design space exploration over a search space consisting of all possible settings for tunable processor design parameters. There are two main challenges that need to be addressed in this process.

Firstly, the design space exploration methodology, employed to select microarchitecture configurations of processors for IoT objects, must be temporally efficient. Long processor design time leads to long time to market which results in lowered profits \cite{mgArchHeteroSys2016}\cite{ rsIoTNewMCUDesign2015} and shorter product life cycle \cite{rsIoTNewMCUDesign2015}. The IoT market also lacks accepted industry standards so, those who get to the market first have the greatest opportunity to influence those standards \cite{rsIoTNewMCUDesign2015}.

Secondly, the design space exploration methodology must balance processor power consumption with performance, which are conflicting design metrics \cite{jbMultiObjOpt2008}. It is not possible to have optimal solutions for optimization problems with conflicting design metrics. The optimization problem should instead be modeled as an Optimal Production Frontier problem also known as Pareto Efficiency\cite{sbConvOpt2004} problem. Multiple solutions are obtained for such problems where each solution favors one of the conflicting metrics. The design space exploration methodology must intelligently choose the best trade-off solution based on application specific requirements.

In this paper, we propose a temporally efficient design space exploration methodology for determining power and performance optimized microarchitecture configurations of embedded processors used in IoT objects. We use a combination of exhaustive, greedy and one-shot search methods to perform design space exploration.
We verify the effectiveness of our methodology by testing it on a cycle accurate simulator using a large set of standard benchmarks with varying workloads.

The main contributions of our paper are:
\begin{itemize}
\item We propose a temporally efficient design space exploration methodology to find microarchitecture configurations for low-power and high-performance optimized embedded processors used in IoT objects.
\item We include a threshold parameter in the design space exploration methodology which can be manipulated by the system designer to control design time based on time to market constraints.
\item We propose exhaustive, greedy and one-shot search algorithms which yield microarchitecture configurations which are 2.23\%-3.69\% of the microarchitecture configurations obtained from fully exhaustive search.
\item We distinguish between different microarchitecture configurations based on the size and type of benchmark used, and, relate them with potential use cases in IoT.
\end{itemize}

The remainder of the paper is organized as follows. In Section \ref{related_work}, we present a review of related work. We describe our design space exploration methodology in Section \ref{methodology} and elaborate on its different phases in Section \ref{phases_of_methodology}. In Section \ref{experimental_setup} we describe the cycle-accurate simulator and benchmarks used to test our methodology. We discuss the results in Section \ref{results} and present our conclusions and future research directions in Section \ref{conclusion_and_future_work}.
%
%
\vspace{-4mm}
\section{Related Work}
\label{related_work}
Several SoC design companies have released articles on techniques of increasing processing capabilities in IoT objects. Some articles guide the selection of processors for IoT objects while others describe low power optimized processor architectures for IoT deployments.

ARM proposed a processor architecture consisting of multiple homogeneous processors in a single IoT object each serving a different purpose \cite{armFlexIoTNodes2015}. They defined a system with three Cortex-M processors, one to handle network connectivity, one to manage interface with sensors and actuators and one as a host processor controlling the other two. They stated that multiple processors are better for lowering power consumption in IoT objects since only the processor serving the current task would be in active mode while the rest would be in sleep mode. ARM also proposed a guide to selecting microcontrollers for IoT objects \cite{armIoTSysDesign2015}. In this guide, they argued that high-end microcontrollers were suitable for IoT deployments for two reasons. Firstly, high-end microcontrollers complete processing tasks sooner and can enter sleep mode to conserve power and secondly, larger flash and RAM sizes available with high-end microcontrollers facilitate implementation of complex networking protocols without addition of any new processors in the system. These articles clearly demonstrate the need for having more power-optimized performance in IoT deployments.

Synopsys also proposed the use of multiple processors in IoT deployments \cite{synUltraLowEmbSubsys2014}. They described the use of two-tiered processor architecture in IoT objects -- ultra low power embedded processors used to interface with sensing elements to collect, filter and process data and host processor used to manage embedded processors. Their processor architecture lowered power consumption by keeping power hungry host processor mostly in sleep mode, similar to the concept used by ARM. Synopsys also discussed optimization of processors using configurable hardware extensions for sensor applications \cite{synUltraLowEmbSubsys2014}. They stated that adding custom hardware extensions for executing typical sensor functions reduces the processor cycle count required to execute sensor applications. The reduction in cycle count lowers energy consumption either by lowering the clock frequency and keeping the same execution time, or having the same power but shorter execution time.

Apart from research carried out by SoC design companies, processor design has also been extensively studied in academia \cite{taMicroprocessorArchIoT2015}\cite{jmPowerPerformancePareto2014}. There are many research works in literature involving optimized processor design. Most works employ design space exploration \cite{tcMicroProcDSE2014}\cite{mmDSEforMulticoreArch2008} techniques utilizing search methods like exhaustive and greedy search and optimizing algorithms like genetic and evolutionary algorithms. Givargis et~al. \cite{tgPLATUNE2002} developed an exploration methodology named PLATUNE (PLATform TUNEr) that carried out exhaustive searches in two stages: first, over clusters of strongly interconnected parameters to obtain Pareto-optimal configurations local to each cluster, and second, over all the clusters to obtain a global Pareto-optimal solution. The approach could explore design spaces as large as $10^{14}$ configurations, but it took an order of 1-3 days to complete. Palesi et~al. \cite{mpDSEGA2002} argued that the high exploration time for PLATUNE was due to the formation of large partial search spaces in the clustering process. Palesi et~al. improved the PLATUNE exploration methodology by introducing a new threshold value that distinguished between clusters based on the size of their partial search-space. Exhaustive search method was used for clusters with partial search-spaces smaller than the threshold value and a genetic exploration algorithm was used for larger spaces. Through this improvement, they were able to achieve 80\% reduction in simulation time while still remaining within 1\% of the results obtained from exhaustive search. Genetic algorithms were also used in the system MULTICUBE, by Silvano et~al. \cite{csMulticube2010}. The MULTICUBE system defined an automatic design space exploration algorithm that could quickly determine an approximate Pareto front for a given design requirements.

Munir et~al. \cite{amWSN2013} proposed another alternative to overcome the overhead of exhaustive search in their work on dynamic optimization of wireless sensor networks. Their approach was divided into two phases. In the first phase, a one-shot search algorithm selected initial parameter settings and further ordered the parameters based on their significance towards the application requirements. In the second phase, a greedy algorithm was used to search the design space. Their approach yielded a design configuration that was within 8\% of the optimal configuration while only exploring 1\% of the design space.

In this paper, we improve on the work carried out by Munir et~al. \cite{amWSN2013}. We leverage a similar approach to design space exploration but add two new phases: a set-partitioning phase and an exhaustive search phase. The addition of the exhaustive search phase aims at increasing the degree of closeness to the optimal solution by exploring a larger portion of the design space, as argued by Silvano et~al. \cite{csMulticube2010}. The limit on the number of configurations considered in the exhaustive search is determined by the set-partitioning phase that uses a threshold value \cite{mpDSEGA2002}.%
%
\vspace{-4mm}
\section{Methodology}
\label{methodology}
\begin{figure}[!t]
\centering
\includegraphics[width = 3.5in, bb = 8 7 1343 1695]{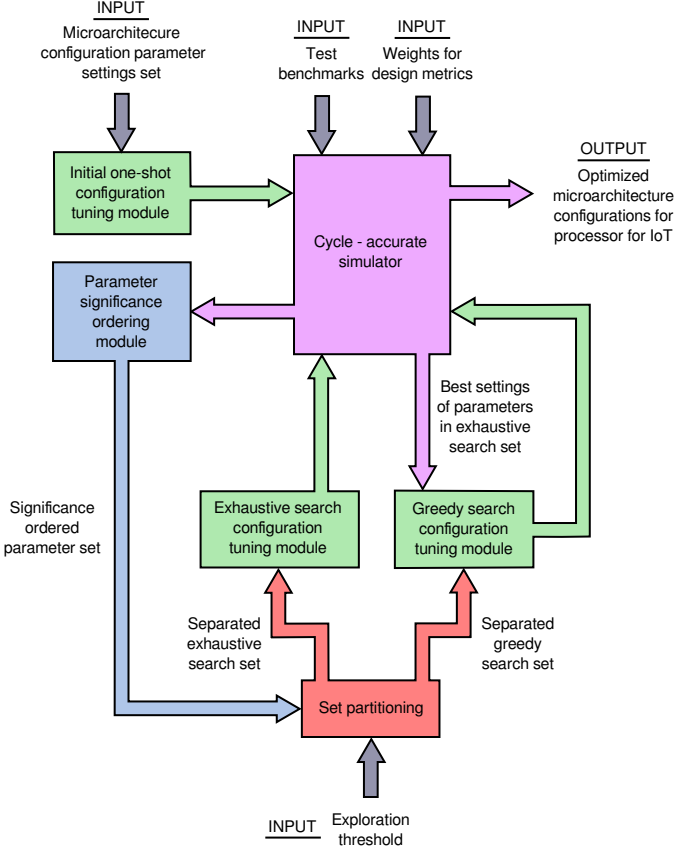}
\caption{Design space exploration methodology for determining optimal microarchitecture configuration of embedded processor for IoT}
\label{figure_methodology}
\end{figure}
Our design space exploration methodology for determining optimal microarchitecture configuration of embedded processors for IoT is shown in Figure \ref{figure_methodology}. Our methodology is implemented in four phases -- initial one-shot search configuration tuning and parameter significance, set-partitioning, exhaustive search configuration tuning and greedy search configuration tuning.

The initial one-shot search configuration tuning and parameter significance phase is carried out by the initial one-shot search configuration tuning module and the parameter significance ordering module. The microarchitecture configuration parameter settings set, which consists of all the possible settings for each tunable microarchitecture parameter, is provided as input to the initial one-shot search configuration tuning module by the system designer. This module uses the parameter settings set to generate initial test configurations. Each initial configuration is passed to a cycle-accurate simulator. The test benchmarks for evaluating the microarchitecture configurations are provided as input to the simulator by the system designer. The simulator executes each initial test configuration separately for each test benchmark specified. The test benchmarks provide varying workloads for testing the initial test configurations. The system designer also provides the weights for balancing design metrics as input to the simulator. These weights are used to specify the preferred tradeoff between conflicting design metrics.

The simulator module evaluates the initial test configurations supplied by the initial one-shot search configuration tuning module to determine the best initial setting for each tunable microarchitecture parameter. The simulation results are forwarded to the parameter significance ordering module where the tunable microarchitecture parameters are ordered based on their significance to the design metrics considered.

The ordered set of significance values is communicated to the set-partitioning module which separates the parameters into two search sets -- exhaustive and greedy. The parameters are separated based on an exploration threshold value provided by the system designer. The exploration threshold value is used to control search space for the exhaustive search phase of our design space exploration methodology. The exhaustive search phase is the longest phase in the design space exploration methodology and processor design time can be significantly altered by varying this exploration threshold value.

The microarchitecture parameters separated out in the exhaustive search set are communicated to the exhaustive search configuration tuning module. This module generates test configurations using all possible combinations of tunable processor design parameters. The parameters which are not in the exhaustive search set retain their best settings from the initial one-shot search configuration tuning process. These test configurations are evaluated on the cycle-accurate simulator to determine a test configuration possessing the best tradeoff between the conflicting design metrics considered. The best settings for the microarchitecture parameters in the exhaustive search set are then communicated to the greedy search configuration tuning module.

The greedy search configuration tuning module generates test configurations using the processor design parameters separated out in the greedy search set. A greedy search algorithm (refer Section \ref{algorithm_phaseIV}) is used to generate these test configurations. The microarchitecture parameters in the exhaustive search set retain their best setting obtained from the exhaustive search simulation process. The parameters which are in neither of the two search sets, retain their best settings from the initial one-shot search configuration tuning process. The best configuration obtained at the end of the greedy search configuration tuning process is communicated back to the processor designer as the optimal microarchitecture of the processor with the preferred tradeoff between the conflicting design metrics.
%
%
\subsection{Defining the Design Space}
\label{methodology_defining_the_design_space}
Consider $n$ number of tunable parameters are available to describe the microarchitecture configuration of an embedded processor for IoT. Let $P$ be the list of these tunable parameters defined as the following set:
\begin{equation}
	P = \{P_{1}, P_{2}, P_{3}, \cdots , P_{n}\}
\end{equation}
Each tunable parameter $P_{i}$ [where $i \in \{1, 2 \cdots n\}$] in the list $P$ is the set of possible settings for $i^{th}$ parameter. Let $L$ be the set containing the size of the set of possible settings for each parameter in list $P$.
\begin{equation}
	L = \{L_{1}, L_{2}, L_{3}, \cdots , L_{n}\}
\end{equation}
such that,
\begin{equation}
	L_{i} = |P_{i}| \;\; \forall \; i \in {1, 2, \cdots , n}
\end{equation}
where $|P_{i}|$ is the cardinal value of set $P_{i}$.

So, each parameter setting set $P_{i}$ in the list $P$ is defined as follows:
\begin{equation}
	P_{i} = \{P_{i1}, P_{i2}, P_{i3}, \cdots , P_{iL_{i}}\} \;\; \forall \; i \in \{1, 2, \cdots, n\}
\end{equation}
The values in the set $P_{i}$ are arranged in ascending order.

The state space for design space exploration is the collection of all the possible configurations that can be obtained using the $n$ parameters.
\begin{equation}
	S = P_{1} \times P_{2} \times P_{3} \times \cdots \times P_{n}
\end{equation}
Here, $\times$ represents the Cartesian product of lists in $P$. Throughout this paper, we use the term $S$ to denote the state space composed of all $n$ tunable parameters. To maintain generality, when referring to a state space composed of $a$ tunable parameters where $a < n$, we attach a subscript to the term $S$.
\begin{equation}
	S_{a} = P_{1} \times P_{2} \times P_{3} \times \cdots \times P_{a} \;\; \forall \; a < n
\end{equation}
We note that the state space of $a$ tunable parameters does not constitute a complete design configuration and is only used as an intermediate when defining our methodology.

We also reserve the use of $\times$ operator in the following manner:
\begin{equation}
	S_{a} = S_{a} \times P_{i} \;\; \forall \; i \in \{1, 2, \cdots, n\}
\end{equation}
This represents the extension of the state space $S_{a}$ to include one new set of parameter settings $P_{i}$ from the list $P$. This operation increases the number of tunable parameters in state space $a$ by one.

When referring to a design configuration that belongs to the state space $S$, we use the term $s$. We attach subscripts to $s$ to refer to specific design configurations. For example, a state $s_{f}$ that consists of the first setting of each tunable parameter can be written as:
\begin{equation}
	s_{f} = (P_{11}, P_{21}, P_{31}, \cdots, P_{n1})
\end{equation}
Similarly, to denote an incomplete/partial design configuration of $a$ tunable parameters we use the term $\delta s_{a}$.
%
%
\subsection{Benchmarks}
\label{methodology_benchmarks}
Each of the configurations, selected from the state space $S$ by our methodology, is tested on $m$ number of test benchmarks. The design metrics for each simulated configuration is collected separately for each benchmark.
%
%
\subsection{Objective Function}
\label{methodology_objective_function}
In our methodology, design configurations are compared with each other based on their objective functions. The objective function of a design configuration is the weighted sum of the normalized design metrics obtained after simulating that design configuration. Let $o$ be the number of design metrics and $V$ be the set of normalized values of design metrics which are obtained from the simulation.
\begin{equation}
	V_{s}^k = \{V_{s1}^k, V_{s2}^k, V_{s3}^k, \cdots , V_{so}^k\} \;\; \forall \; k = {1, 2, \cdots , m}
\end{equation}
Let $w$ be the set of weights for the design metrics based on the requirements of the targeted application. These weights are set by the system designer.
\begin{equation}
	w = \{w_{1}, w_{2}, w_{3}, \cdots , w_{o}\}\
\end{equation}
such that,
\begin{equation}
	0 \leq w_{l} \leq 1\ \forall \; l = {1, 2, \cdots , o}
\end{equation}
and,
\begin{equation}
	\sum w_{l} = 1 \;\; \forall \; l = {1, 2, \cdots , o}
\end{equation}
The objective function $\mathcal{F}$ of a design configuration $s$ for a test benchmark $k$ is defined as follows:
\begin{equation}
	\mathcal{F}_{s}^k = \sum w_{l}V_{sl}^k \;\; \forall \; l = {1, 2, \cdots , o}
\end{equation}
The optimization problem, considered in this paper, is to minimize the value of the objective function $\mathcal{F}$. The design metrics are chosen such that the minimization of their values is the favorable design choice. For example, when considering the performance metric, the design goal is to maximize performance. To model this into the objective function which we use execution time to measure performance. Minimizing execution time would fit with minimizing the objective function while still modeling the design goal of maximizing performance. The optimization problem for each test benchmark $k$ is defined as follows:
\begin{equation}
\begin{split}
	min. \;\; & F_{s}^k \\
	s.t.\;\; & s \in S
\end{split}
\end{equation}

Table \ref{Table_List_Symbols} presents the symbols established in this section in list form.

\begin{table}[!t]
\caption{List of symbols}
\label{Table_List_Symbols}
\centering
\begin{tabular}{| c | p{2.75in} |}\hline
\textbf{Symbol} & \textbf{Description }\\\hline
n & Number of tunable microarchitecture parameters \\\hline
$P$ & List of tunable microarchitecture parameters \\\hline
$P_{i}$ & Set of possible settings for $i^{th}$ tunable microarchitecture parameter \\\hline
$L$ & Size of set of possible settings for each tunable microarchitecture parameter \\\hline
$L_{i}$ & Cardinal value of set $P_{i}$ \\\hline
$S$ & State space for design space exploration \\\hline
$S_{a}$ & Partial/Incomplete state space \\\hline
$s_{tag}$ & State in state space $S$ with `tag' identifier \\\hline
$\delta s_{a}$ & State in partial state space $S_{a}$ \\\hline
$m$ & Number of test benchmarks \\\hline
$o$ & Number of design metrics \\\hline
$V_{s}^{k}$ & Set of normalized values obtained for design metrics from simulation of state $s$ for $k^{th}$ benchmark \\\hline
$w$ & Set of weights for design metrics \\\hline
$w_{l}$ & Weight for $l^{th}$ design metric \\\hline
$\mathcal{F}_{s}^{k}$ & Objective function obtained from simulating state $s$ for $k^{th}$ benchmark \\\hline

\end{tabular}
\end{table}

%
%
\section{Phases of Methodology}
\label{phases_of_methodology}

Our proposed design space exploration methodology consists of four distinct phases. In this section, we elaborate on the steps involved in each phase using the notation set up in Section \ref{methodology}.
%
%
\subsection{Phase I : Initial One-Shot Search Configuration Tuning and Parameter Significance}
\label{algorithm_phaseI}

In this phase of our methodology, best initial setting for each tunable microarchitecture parameter in set $P$ is determined by using a one-shot search configuration tuning process. The one-shot search process is based on single factor analysis which is an effective heuristic approach used in design space exploration \cite{dsDSEUDOE11}. Unlike single factor analysis wherein parameters can have only two settings, a zero value and a non-zero value setting, one-shot search works on parameters with more than two non-zero value settings. In one-shot search process, parameters are evaluated on a one by one basis. Two test configurations are generated for each parameter, one with the first setting and one with the last settings from the list of settings for the current parameter. The remaining parameters are arbitrarily set to their first setting from their corresponding list of settings.

\begin{algorithm}
\caption{Initial One-Shot Search Configuration Tuning and Parameter Significance}
\label{algorithm_parameter_significance}

\LinesNumbered
\DontPrintSemicolon
\SetKwData{Left}{left}\SetKwData{This}{this}\SetKwData{Up}{up}
\SetKwFunction{Union}{Union}\SetKwFunction{FindCompress}{FindCompress}
\SetKwInOut{Input}{input}\SetKwInOut{Output}{output}
\KwIn{$P$ - List of Tunable Parameters}
\KwOut{$B$ - Set of Best Settings; $D$ - Significance of Parameters with respect to Objective Function}

\BlankLine
\For {$i \leftarrow 1$ \KwTo $n$}
    {
    $s_{f} = \{P_{i1}\}$\;
    $s_{l} = \{P_{iL[i]}\}$\;
    \For {$j \leftarrow 1$ \KwTo $n$}
    {
    \If {$i \neq j$}
        {
        $s_{f} = s_{f} \cup \{P_{j1}\}$\;
        $s_{l} = s_{l} \cup \{P_{j1}\}$\;
        }
    }
    \For {$k \leftarrow 1$ \KwTo $m$}
    {
    Explore $k^{th}$ benchmark using configuration $s_{f}$\;
    \ \ \ \ Calculate $\mathcal{F}_{s_{f}}^{k}$\;
    Explore $k^{th}$ benchmark using configuration $s_{l}$\;
    \ \ \ \ Calculate $\mathcal{F}_{s_{l}}^{k}$\;
    $D^{k}_{i}$ = $\mathcal{F}^{k}_{l} - \mathcal{F}^{k}_{f}$\;
    \label{algorithm_parameter_significance_ref1}
    \eIf {$D^{k}_{i} > 0$}
        {
        $B^{k}_{i} = P_{i1}$\;
        \label{algorithm_parameter_significance_ref2}
        }
        {
        $B^{k}_{i} = P_{iL[i]}$\;
        \label{algorithm_parameter_significance_ref3}
        }
    }
}
\end{algorithm}

The steps involved in initial one-shot search configuration tuning and determining parameter significance are detailed in Algorithm \ref{algorithm_parameter_significance}. The first and last test configurations generated for evaluating a tunable microarchitecture parameter, $P_{i}$ in set $P$, are denoted by $s_{f}$ and $s_{l}$, respectively. These configurations are tested on the cycle-accurate simulator. From the results of the simulation, objective functions, $\mathcal{F}_{s_{f}}$ and $\mathcal{F}_{s_{l}}$ corresponding to $s_{f}$ and $s_{l}$, respectively, are determined. The objective function values are used to determine best initial setting as well as significance of each microarchitecture parameter. The magnitude of the difference between $\mathcal{F}_{s_{f}}$ and $\mathcal{F}_{s_{l}}$, which is stored in parameter significance set $D$ (line \ref{algorithm_parameter_significance_ref1}), is used as parameter significance. The higher the magnitude of a difference $D_{i}^{k}, \; i \in \{1, 2, 3, \ldots, n\}$ for a benchmark $k, \; k \in \{1, 2, 3, \ldots, m\}$, the higher is the significance of parameter $P_{i}$ to the workload characterized by benchmark $k$. The sign of the difference between $\mathcal{F}_{s_{f}}$ and $\mathcal{F}_{s_{l}}$ is used to pick the best initial setting for parameter $P{i}$. If the difference is positive, then the first setting of parameter $P_{i}$ is chosen as the best setting, otherwise the last setting is chosen. The best settings for the parameters are stored in the set of best settings $B_{i}^{k}$ (lines \ref{algorithm_parameter_significance_ref2} and \ref{algorithm_parameter_significance_ref3}).
%
%
\subsection{Phase II : Set-Partitioning}
\label{algorithm_phaseII}

\begin{algorithm}
\caption{Set-Partitioning}
\label{algorithm_set_partitioning}

\LinesNumbered
\DontPrintSemicolon
\SetKwData{Left}{left}\SetKwData{This}{this}\SetKwData{Up}{up}
\SetKwFunction{Union}{Union}\SetKwFunction{FindCompress}{FindCompress}
\SetKwInOut{Input}{input}\SetKwInOut{Output}{output}
\KwIn{$D$ - Significance of Parameters towards Objective Function; $I$ - Index Set; $T$ - Exhaustive Search Threshold Factor}
\KwOut{$\mathcal{E}$ - Set of Parameters for Exhaustive Search; $\mathcal{G}$ - Set of Parameters for Greedy Search}
\BlankLine
$\mathcal{E} = \emptyset\ and\ \mathcal{G} = \emptyset$\;
\For {$k \leftarrow 1$ \KwTo $m$}
    {
    sortDescending ($\mid D^{k} \mid$)- s.t. index information of the sorted values is preserved in $I^{k}$\;
    \label{algorithm_set_partitioning_ref1}
    sort($P^{k}$) and sort($L^{k}$) w.r.t. index information in $I^{k}$\;
    $num_{\mathcal{E}} = 1$ and $i = 1$\;
    \While {$num_{\mathcal{E}} \leq T$}
        {
		\label{algorithm_set_partitioning_ref2}
        $num_{\mathcal{E}} = num_{\mathcal{E}} \times L_{i}^{k}$\;
	    \eIf {$num_{\mathcal{E}} \leq T$}
	    {
           $\mathcal{E}^{k} = \mathcal{E}^{k} \cup \{P_{i}\}$\;
           $i = i + 1$\;
	    }
        {
	      \textbf{break}\;
        }
        }
    $num_{\mathcal{G}} = ceil((|P^{k}| - |\mathcal{E}^{k}|)\ /\ 2)$\;
    \label{algorithm_set_partitioning_ref3}
    \While {$num_{\mathcal{G}} > 0$}
        {
        $\mathcal{G}^{k} = \mathcal{G}^{k} \cup \{P_{i}^{k}\}$\;
        $num_{\mathcal{G}} = num_{\mathcal{G}} - 1$\;
        $i = i + 1$\;
        }
    }
\end{algorithm}

The set-partitioning phase, presented in Algorithm \ref{algorithm_set_partitioning}, shows how the parameter significance values determined in the first phase of our methodology are used to separate the list of tunable microarchitecture parameters into exhaustive and greedy search sets. First, the parameter significance set $|D^{k}|$ for each benchmark $k, \; k \in \{1, 2, 3, \ldots, m\}$, is sorted in descending order of magnitude using the sortDescending($|D^{k}|$) function. The index information of the sorted values is preserved in a set of indexes $I^{k}$ (line \ref{algorithm_set_partitioning_ref1}). For example, if the fifth entry $D_{5}^{k}$ has the greatest value, $D_{5}^{k}$ will become the first entry in the set $D^{k}$ and first entry in the set of indexes $I^{k}$ will be 5, that is, $I_{1}^{k} = 5$. The set of indexes, $I^{k}$, is used to sort the list of tunable microarchitecture parameters, $P^{k}$, and list of set sizes, $L^{k}$. After sorting, the parameters with higher significance lie towards the start of the set and the parameters with lower significance lie towards the end of the set. The list of parameters is then divided into three subsets, exhaustive search, greedy search and one-shot search sets. The exhaustive search set gets parameters with the highest significance. The number of parameters separated into the exhaustive search set depends on the exploration threshold value, $T$, provided by the system designer. The threshold value $T$ limits the size of the partial search space of the exhaustive search set, $num_{\mathcal{E}}$ (line \ref{algorithm_set_partitioning_ref2}).

After separating out exhaustive search set, the parameters remaining in the parameter list are separated into greedy search and one-shot search sets. The list of remaining parameter is divided into two halves (line \ref{algorithm_set_partitioning_ref3}) and the upper half $ceil((|P| - |\mathcal{E}^{k}|)/2)$ is separated as the greedy search set and the lower half is separated as one-shot search set. We observe empirically that dividing the list of remaining parameters into halves provides efficient design space exploration without significantly compromising the solution quality. The parameters separated as one-shot search set are not explored further and are left at the best settings determined for them in Algorithm \ref{algorithm_parameter_significance}.
%
%
\subsection{Phase III : Exhaustive Search Configuration Tuning}
\label{algorithm_phaseIII}

\begin{algorithm}
\caption{Exhaustive Search}
\label{algorithm_exhaustive_search}

\LinesNumbered
\DontPrintSemicolon
\SetKwData{Left}{left}\SetKwData{This}{this}\SetKwData{Up}{up}
\SetKwFunction{Union}{Union}\SetKwFunction{FindCompress}{FindCompress}
\SetKwInOut{Input}{input}\SetKwInOut{Output}{output}
\KwIn{$P$ - List of Tunable Parameters; $B$ - Set of Best Settings for One-shot Search; $\mathcal{E}$ - List of Parameters for Exhaustive Search}
\KwOut{$B$ - List of Best Settings for One-shot and Exhaustive Search}
\BlankLine
$s_{\mathcal{E}} = \emptyset$\;
$\delta s_{\mathcal{E}} = \emptyset\ and\ \delta s_{\mathcal{E}^{'}} = \emptyset$\;

\For {$k \leftarrow 1$ \KwTo $m$}
    {
    $\mathcal{F}^{k}_{s_{b}} = \infty$\;
    \For {$i \leftarrow 1$ \KwTo $n$}
        {
        \If {$P_{i} \notin \mathcal{E}^{k}$}
        		{
	        $\delta s^{k}_{\mathcal{E}^{'}} = \delta s^{k}_{\mathcal{E}^{'}} \cup \{B^{k}_{i}\}$\;	
	        \label{algorithm_exhaustive_search_ref1}
        		}
        }

    \For {$i \leftarrow 1$ \KwTo $n$}
        {
        \If {$P_{i} \in \mathcal{E}^{k}$}
	        {
        		$S^{k}_{\mathcal{E}} = S^{k}_{\mathcal{E}} \times P_{i}$\;
        		\label{algorithm_exhaustive_search_ref2}
        		}
        	}

    \For {$j \leftarrow 1$ \KwTo $|S^{k}_{\mathcal{E}}|$}
        {
        	$\delta useds^{k}_{\mathcal{E}j}$ is a partial configuration in state space $S^{k}_{\mathcal{E}}$\;
        	\label{algorithm_exhaustive_search_ref3}
        $s^{k}_{\mathcal{E}} = \delta s^{k}_{\mathcal{E}j} \cup \delta s^{k}_{\mathcal{E}^{'}}$\;
        \label{algorithm_exhaustive_search_ref4}
        Explore $k^{th}$ benchmark using configuration $s^{k}_{\mathcal{E}}$\;
        \ \ \ \ Calculate $\mathcal{F}^{k}_{\mathcal{E}}$\;
        \If {$\mathcal{F}^{k}_{s_{\mathcal{E}}} < \mathcal{F}^{k}_{s_{b}}$}
            {
            \label{algorithm_exhaustive_search_ref5}
            $\mathcal{F}^{k}_{s_{b}} = \mathcal{F}^{k}_{s_{\mathcal{E}}}$\;
            $B^{k} = s^{k}_{\mathcal{E}}$\;
            }
        }
    }
\end{algorithm}

Algorithm \ref{algorithm_exhaustive_search} details the steps involved in the exhaustive search process. The exhaustive search process determines the best settings for the parameters in the exhaustive search set $\mathcal{E}$. First, the settings for the parameters that are not in the exhaustive search set $\mathcal{E}$ are assigned (line \ref{algorithm_exhaustive_search_ref1}). These parameters are assigned their best settings from the set of best settings $B_{i}^{k}$ as determined in the initial one-shot search configuration tuning process described in Algorithm \ref{algorithm_parameter_significance}. These settings make up the partial test design configuration $\delta s_{\mathcal{E}^{'}}$. Next, a partial state space $S_{\mathcal{E}}$ is formed for the parameters in the exhaustive search set $\mathcal{E}$ (line \ref{algorithm_exhaustive_search_ref2}). Every possible partial test design configuration, $\delta s_{\mathcal{E}j}$ (line \ref{algorithm_exhaustive_search_ref3}), in the partial state space $S_{\mathcal{E}}$, is combined with the partial test design configuration $\delta s_{\mathcal{E}^{'}}$ to form complete simulatable test design configurations. Each complete test design configuration is evaluated on the simulator. An objective function value, $\mathcal{F}_{s_{\mathcal{E}}}$, is obtained for each complete test design configuration, $s_{\mathcal{E}}$, from the simulator. The algorithm keeps track of the smallest objective function value encountered in the search process in $\mathcal{F}_{s_{b}}$ which represents the best objective function value. When a design configuration results in an objective function that has a value less than $\mathcal{F}_{s_{b}}$ (line \ref{algorithm_exhaustive_search_ref5}), then $\mathcal{F}_{s_{b}}$ is changed to the new minimum value and the set of best settings $B$ is updated with the corresponding design configuration.
%
%
\subsection{Phase IV : Greedy Search Configuration Tuning}
\label{algorithm_phaseIV}

In the final phase of our methodology, described in Algorithm \ref{algorithm_greedy_search}, the best settings for the parameters in greedy search set $\mathcal{G}$ are determined. For each parameter in the set $\mathcal{G}$, the sign of the parameter significance is checked to determine whether the first setting or last setting was chosen as the best setting in the first phase of our methodology. If the sign of parameter significance is positive, then it indicates that first setting for that parameter yields a smaller objective function as compared to the last. If the sign is negative then it indicates that the last setting for that parameter yields a smaller objective function as compared to the first. We assume that the setting that yields the smallest objective function lies closer towards the setting that yields the smallest objective function in the initial one-shot search configuration tuning process. To ensure that the search process starts from the setting that yielded the smallest objective function in the initial one-shot search configuration tuning process, we sort the set of parameter settings $P_{i}$ in descending order (for last setting as best setting) or left unchanged in default ascending order (for first setting as best setting) (line \ref{algorithm_greedy_search_ref1}).

\begin{algorithm}
\caption{Greedy Search}
\label{algorithm_greedy_search}

\LinesNumbered
\DontPrintSemicolon
\SetKwData{Left}{left}\SetKwData{This}{this}\SetKwData{Up}{up}
\SetKwFunction{Union}{Union}\SetKwFunction{FindCompress}{FindCompress}
\SetKwInOut{Input}{input}\SetKwInOut{Output}{output}
\KwIn{
$P$ - List of Tunable Parameters, $D$ - Significance of Parameters towards Objective Function, $B$ - Set of Best Settings for One-shot and Exhaustive Search, $\mathcal{E}$ - Set of Parameters for Exhaustive Search, $\mathcal{G}$ - Set of Parameters for Greedy Search}
\KwOut{$B$ - Complete set of Best Settings}
\BlankLine
$s_{\mathcal{G}} = \emptyset$\;
$\delta s_{\mathcal{G}^{'}} = \emptyset$\;
$\mathcal{G}_{P} = \emptyset$\;
\For {$k = 1$ \KwTo $m$}
    {
    $\mathcal{F}^{k}_{s_{b}} = \infty$\;
    \For {$i \leftarrow 1$ \KwTo $n$}
        {
        \If {$P_{i} \in \mathcal{G}^{k}$}
            {
            \If {$D^{k}_{i} < 0$}
                {
                \label{algorithm_greedy_search_ref1}
                $\mathcal{G}_{P}$ = sortDescending ($P_{i}$)\;
                }
            \For {$j \leftarrow 1$ \KwTo $n$}
                {
                \If { $P_{j} \neq \mathcal{G}_{P}$}
                    {
                    \label{algorithm_greedy_search_ref2}
                    $\delta s^{k}_{\mathcal{G}_{P}^{'}} = \delta s^{k}_{\mathcal{G}_{P}^{'}} \cup \{B^{k}_{j}\}$\;
                    }
                }
            \For {$l \leftarrow 1$ \KwTo $L_{i}$}
                {
                $s^{k}_{\mathcal{G}} = \delta s^{k}_{\mathcal{G_{P}}^{'}} \cup \{\mathcal{G}_{Pl}\}$\;
                \label{algorithm_greedy_search_ref3}
                Explore $k^{th}$ benchmark using configuration $s^{k}_{\mathcal{G}}$\;
                \ \ \ \ Calculate $\mathcal{F}^{k}_{s_{\mathcal{G}}}$\;
                \eIf {$\mathcal{F}^{k}_{s_{\mathcal{G}}} < \mathcal{F}^{k}_{s_{b}}$}
                    {
                    \label{algorithm_greedy_search_ref4}
                    $\mathcal{F}^{k}_{s_{b}} = \mathcal{F}^{k}_{s_{\mathcal{G}}}$\;
                    $B^{k}_{i} = \mathcal{G}_{Pj}$\;
                    }
                    {
                    \textbf{break}\;
                    }
                }
            }
        }
    }
\end{algorithm}

In the greedy search process, the parameters in the greedy search set are considered one at a time. First, a partial test design configuration $\delta s_{\mathcal{G_{P}}^{'}}$ is formed using the exhaustive search set, the one-shot search set and the non-current parameters in greedy search set. The parameters in the exhaustive search set, $\mathcal{E}$, are assigned their best values as determined in the exhaustive search configuration tuning process. The parameters in the one-shot search set retain the best settings determined in the initial one-shot configuration tuning process. The non-current parameters in the greedy search set, $\mathcal{G}$, are assigned best settings in one of two ways. If the non-current parameter has already been processed by the greedy search optimization process, then the parameter is assigned the best setting obtained from that process. If the non-current parameter has not been processed yet, then the parameter is assigned the best setting obtained from the initial one-shot search configuration tuning process.

The partial test design configuration $\delta s_{\mathcal{G_{P}}^{'}}$ is then combined with the settings for the current parameter being processed to form the complete simulatable test design configuration $s_{\mathcal{G}}$ (line \ref{algorithm_greedy_search_ref3}). This configuration is evaluated on the cycle-accurate simulator. The resulting objective function, $\mathcal{F}_{s_{\mathcal{G}}}$, is compared with the best objective function $\mathcal{F}_{s_{b}}$, which holds the smallest value objective function encountered thus far in the search process. Similar to the exhaustive search process, when a design configuration results in an objective function that has a value less than $\mathcal{F}_{s_{b}}$ (line \ref{algorithm_greedy_search_ref4}), then $\mathcal{F}_{s_{b}}$ is changed to the new minimum value and the set of best settings $B_{i}^{k}$ is updated with the corresponding design configuration. However, when the search process encounters a design configuration that results in an objective function that has a value greater than $\mathcal{F}_{s_{b}}$, then the search process for the current parameter is terminated and the next parameter in the parameter list $\mathcal{G}$ is explored.
%
%
\section{Experimental Setup}
\label{experimental_setup}
We used the ESESC\cite{ekaESESC2013} (Enhanced Super EScalar) simulator to simulate all the test microarchitecture configurations generated by our methodology. The ESESC simulator is a fast cycle-accurate chip multiprocessor simulator. It models an out-of-order RISC (Reduced Instruction Set Computing) processor running ARM instruction set.

We used benchmarks from the PARSEC and SPLASH2\cite{cbPARSEC2011, ybPARSEC3_TUT2011} benchmark suite to test our methodology. The PARSEC and SPLASH2 benchmark suite is a collection of standardized benchmarks which provides a diverse range of workloads for evaluation of processors.

We used the following benchmarks from the PARSEC and SPLASH2 suite to test our methodology.

\textbf{PARSEC Benchmarks}: Blackscholes, Canneal, Facesim, Fluidanimate, Freqmine, x264

\textbf{SPLASH2 Benchmarks}: Cholesky, FFT, LU\_cb, LU\_ncb, Ocean\_cp, Ocean\_ncp, Radiosity, Radix, Raytrace

The methodology phases were implemented using PERL\cite{pmPERL2015}. The results from the simulation processes were collected in MS Excel using Excel-Writer-XLSX\cite{jmExcel2015} tool for PERL.

We tested our design space exploration methodology separately for low-power and high-performance processor design. We combined the microarchitecture configurations obtained from these tests to form a two-tiered heterogeneous processor architecture. The microarchitecture configuration obtained from the low-power processor design tests were used to implement the low-power optimized interface processors, the lower tier of the two-tiered architecture. The microarchitecture configuration obtained from the high-performance processor design tests were used to implement the high-performance optimized host processor, the upper tier of the two-tiered architecture.

\begin{table}[!h]
\caption{Mircoarchitecture configuration parameter settings set}
\label{table_parameter_settings}
\centering
\begin{tabular}{| l | c | c |}\hline
\multirow{2}{*}{\textbf{Parameter Name}}          & \multicolumn{2}{ c |}{\textbf{Set of Settings}} \\\cline{2-3}
                                 & \textbf{Low-Power}   & \textbf{High-Performance} \\\hline
Cores                            & 1, 2, 4            & 2, 4, 8 \\\hline
Frequency (MHz)                  & 75, 100, 125, 150  & 1700, 2200, 2800, 3200 \\\hline
L1-I Cache Size (kB)         & 8, 16, 32, 64      & 8, 16, 32, 64, 128 \\\hline
L1-D Cache Size (kB)         & 8, 16, 32, 64      & 8, 16, 32, 64, 128 \\\hline
L2 Cache Size (kB)           & 256, 512, 1024     & 256, 512, 1024 \\\hline
L3 Cache Size (kB)           & 2048, 4096         & 2048, 4096, 8192 \\\hline
\end{tabular}
\end{table}

The list of microarchitecture parameters considered for testing our methodology along with the set of possible settings for each parameter is listed in Table \ref{table_parameter_settings}. We used different range of settings for low-power and high-performance processor design. The range of settings listed in Table \ref{table_parameter_settings} under low-power design were used for the design of low-power optimized interface processors. The design space cardinality for low-power processor design was 1,152 configurations. The range of settings listed in Table \ref{table_parameter_settings} under high-performance design were used for the design of high-performance optimized host processor. The design space cardinality for high-performance processor design was 2,700 configurations.

\begin{table}[!t]
\centering
\caption{Weights for design metrics}
\label{table_design_metric_weights}
\begin{tabular}{| l | c | c |}\hline
\textbf{Configuration}     & \textbf{Power} & \textbf{Performance}  \\\hline
Low-Power         & 0.9   & 0.1          \\\hline
High-Performance  & 0.1   & 0.9          \\\hline
\end{tabular}
\end{table}

We used power and performance as design metrics to evaluate the microarchitecture configurations for both low-power and high-performance optimized processors. We used normalized value of total dynamic power and leakage power \cite{aflISPCMES2015} across all the cores in the processor as the power metric and the normalized value of total execution time as the performance metric. We used the weights presented in Table \ref{table_design_metric_weights} to specify the preference for the conflicting design metrics of power and performance. The linear objective function used for the evaluation of the test microarchitecture configurations was:

\begin{equation}
\mathcal{F} = w_{\mathcal{P}} \cdot \mathcal{P} + w_{E} \cdot E
\label{equation_objective_function}
\end{equation}

where,
\begin{equation}
\begin{aligned}
\mathcal{P} &= Dynamic\;Power + Leaked\;Power \\
E &= Total\;Execution\;Time \\
\end{aligned}
\end{equation}
%
%
\section{Results}
\label{results}
In this section, we present the results obtained while testing our methodology. This section is divided into two subsections. In the first subsection, we present results to validate our design space exploration methodology and in the second subsection, we discuss some of the applicability of some of the microarchitecture configurations to important IoT use cases.
\subsection{Evaluation of design space exploration methodology}
\label{results_evaluation_DSE}
For evaluating our methodology, we compared our microarchitecture configuration results with those obtained from a fully exhaustive search of the design space. We tested our methodology with an exploration threshold of $T$ = 150. This threshold value is an upper bound which limits the partial state space for the exhaustive search phase of our methodology.
\subsubsection{Parameter significance}
\begin{figure}[!t]
\centering
\includegraphics[width = 3.5in, bb = 0 0 476 386]{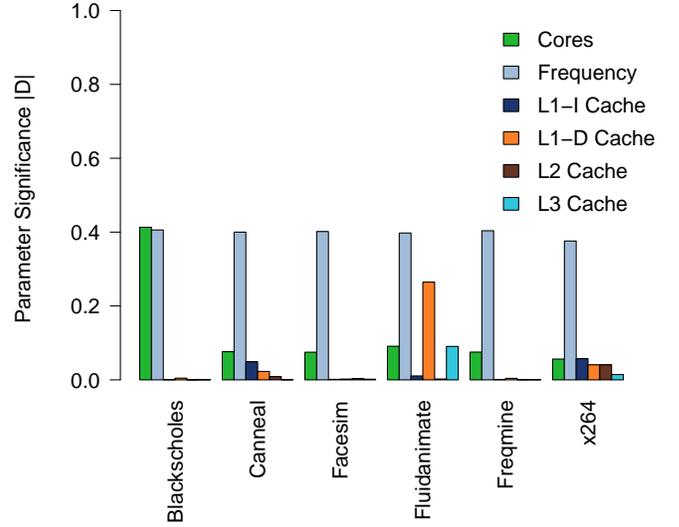}
\caption{Significance of microarchitecture configuration parameters for PARSEC benchmarks for high-performance optimized processor for IoT}
\label{figure_significance_x264}
\end{figure}
Figure \ref{figure_significance_x264} shows the normalized values of parameter significance for different PARSEC benchmarks. The normalization is carried out using the maximum values for total power and total execution time obtained in the initial one-shot search configuration tuning process. The parameter significance values are calculated in the first phase of our methodology, initial one-shot search configuration tuning. We observe that the significance of each of the tunable processor design parameters varies based on the type of workload offered by the test benchmarks. For each of the test benchmarks, there are at most three significant processor design parameters. We note that the operating frequency is the processor design parameter with the highest significance for most of the test benchmarks followed by core count, which is the second most significant design parameter. For certain test benchmarks, the size of the L1-I cache and L1-D cache are also highly significant to overall design. The large significance in cache sizes is a result of large working sets with fine data-parallel granularity offered by those test benchmarks.
\subsubsection{Selecting a favorable tradeoff solution}
\begin{figure}[!t]
\centering
\includegraphics[width = 3.5in, bb = 0 0 706 536]{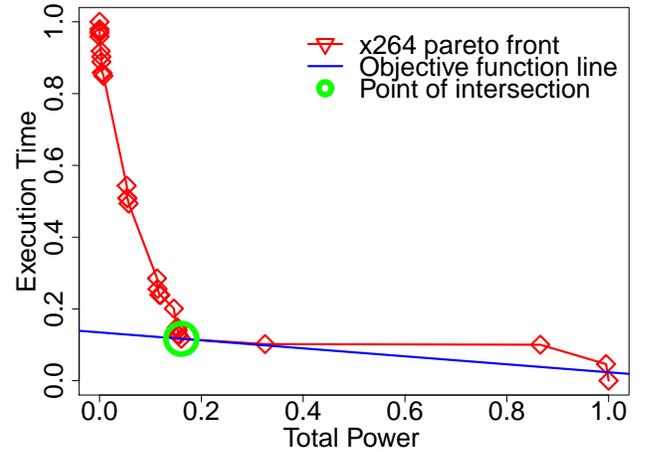}
\vspace{-10mm}
\caption{Linear objective function plotted with Pareto front for x264 (PARSEC) benchmark for high-performance optimized processor for IoT}
\label{figure_pareto_tangent_x264}
\end{figure}

Figure \ref{figure_pareto_tangent_x264} shows the Pareto front obtained for x264 (PARSEC) benchmark for high-performance optimization requirement. The Pareto front is generated using the normalized values of total power and execution time design metrics. The front represents the conflicting interdependency between power and performance in a processor. It shows that increasing the performance of a processor degrades its power efficiency whereas increasing power efficiency degrades performance. It is thus impossible to determine a microarchitecture configuration which results in both these metrics having optimal values. The goal of the design space exploration methodology is to determine a balance between these conflicting design metrics. A suitable tradeoff between these metrics is selected by using the preference specified using the weights assigned to each metric. In our experiments, we specified $w_{\mathcal{P}}$ and $w_{E}$ as the weights for power and performance metrics respectively to define a linear objective function (Equation \ref{equation_objective_function}). Figure \ref{figure_pareto_tangent_x264} shows the objective function plotted along with the Pareto front. We note that the objective function forms a straight line in the power-performance graph with the slope $-w_{\mathcal{P}}/w_{E}$. We observe that the objective function is tangent to the Pareto front at the power-performance value pair of the microarchitecture configuration obtained as solution by our methodology.

\subsubsection{Comparison with fully exhaustive search}
\begin{table}[!t]
\caption{Comparison of mircoarchitecture configurations obtained for x264 (PARSEC) benchmark for high-performance optimized processor for IoT}
\label{table_results_configuration_x264}
\centering
\begin{tabular}{| l | c | c |}\hline
\multirow{3}{*}{\textbf{Parameter Name}}          & \multicolumn{2}{ c |}{\textbf{Microarchitecture Configuration}} \\\cline{2-3}
                                 & Proposed & Fully Exhaustive \\
                                 & Methodology & Search \\\hline
Cores                            & 2 & 2 \\\hline
Frequency (MHz)                  & 3200 & 3200 \\\hline
L1-I Cache Size (kB)         & 64 & 128 \\\hline
L1-D Cache Size (kB)         & 64 & 128 \\\hline
L2 Cache Size (kB)           & 1024 & 256 \\\hline
L3 Cache Size (kB)           & 2048 & 8192 \\\hline
\textbf{Total Power (W)}         & 1.597 & 1.600 \\\hline
\textbf{Execution Time (ms)}     & 35.142 & 34.152 \\\hline
\end{tabular}
\end{table}

We verified the microarchitecture configuration obtained as solution from our methodology by comparing it against the solution obtained by running a fully exhaustive search of the design space. We present a comparison of the x264 (PARSEC) benchmark as an example in Table \ref{table_results_configuration_x264}. The table shows a side-by-side comparison of the microarchitecture configurations obtained from our proposed methodology with the same obtained from fully exhaustive search. Comparing these values, we see that significant parameters like operating frequency and core-count match exactly while other parameters only differ slightly. The table also contains the values of the total power and execution time obtained for both configurations. Comparing the values of these design metrics, we see that the total power and execution time values obtained from our methodology are within -0.18\%  and 2.89\% respectively of the total power and execution time values obtained from fully exhaustive exploration.

Using our methodology, on average we achieve microarchitecture configurations with total power values within 2.23\% for low-power optimized processor and execution time within 3.69\% for high-performance optimized processors as compared to fully exhaustive search. These configurations are obtained by exploring only 3\%--5\% of the processor design space which results in our methodology having an average speedup of 24.16$\times$ as compared to fully exhaustive exploration of the design space.
\subsection{Application scopes in IoT}
\label{results_application_IoT}

Based on the type and size of workload offered by the test benchmarks, we separate them into four different categories each of which relates to an IoT application or process. Table \ref{table_results_benchmark_application_scope} shows the categorization of some of the key test benchmarks. The Cholesky and Radix benchmarks from the SPLASH2 benchmark suite are categorized under data sensing and aggregation. The Cholesky benchmark is a sparse matrix factorization kernel and the Radix benchmark is an integer sort kernel\cite{scwSPLASH2Programs1995}. The Cholesky benchmark is representative of  data sensing in IoT applications, where data is acquired from multiple sensor sources and transformed into a more useful format. The Radix benchmark is representative of data aggregation, where indexing, sorting and storing operations are carried out on sensed data. These benchmarks are useful in determining the microarchitecture configurations of low-power optimized interface processors for the two-tiered heterogeneous processor architecture.

\begin{table}[!t]
\caption{Categorization of test benchmarks according to IoT application}
\label{table_results_benchmark_application_scope}
\centering
\begin{tabular}{| l | l |}\hline
\textbf{IoT Application}             & \textbf{Benchmarks} \\\hline
Data sensing and aggregation         & Cholesky, Radix \\\hline
Data analysis and Data mining        & Blackscholes, Freqmine \\\hline
Graphics                             & Facesim, Fluidanimate \\\hline
Signal processing and Communication  & FFT \\\hline
\end{tabular}
\end{table}

\begin{table}[!t]
\caption{Mircoarchitecture configurations for low-power optimized processors for IoT}
\label{table_results_low_power}
\centering
\begin{tabular}{| l | >{\centering} p{1.75cm} | c |}\hline
\multirow{2}{*}{\textbf{Parameter Name}} & \multicolumn{2}{ c |}{\textbf{Microarchitecture Configuration}} \\\cline{2-3}
                             & \textbf{Cholesky} & \textbf{Radix} \\\hline
Cores                        & 1 & 1 \\\hline
Frequency (MHz)              & 75 & 75 \\\hline
L1-I Cache Size (kB)         & 8 & 8 \\\hline
L1-D Cache Size (kB)         & 32 & 64 \\\hline
L2 Cache Size (kB)           & 256 & 256 \\\hline
L3 Cache Size (kB)           & 2048 & 4096 \\\hline
\textbf{Total Power (W)}     & 0.0934 & 0.0935 \\\hline
\textbf{Execution Time (ms)} & 327.958 & 332.535 \\\hline
\end{tabular}
\end{table}

The remaining categories all model more complex applications requiring high level of processing capabilities. The Blackscholes and Freqmine benchmarks from the PARSEC benchmark suite are listed under data analysis and data mining. The Blackscholes benchmark is a financial analysis benchmark that analytically solves large sets of partial differential equations\cite{cbPARSEC2011}. The Freqmine benchmark is a data mining kernel which implements Frequent Itemset Mining\cite{cbPARSEC2011}. These benchmarks are representative of data analysis and filtering operations that need to be carried out on large volumes of sensor data in an IoT network.

The Facesim and Fluidanimate benchmarks from the PARSEC benchmark suite are listed under graphics. The Facesim benchmark generates a visually realistic model of a human face and the Fluidanimate benchmark simulates an incompressible fluid for interactive animation purposes\cite{cbPARSEC2011}. Graphical applications are important in IoT objects which need to interact with users via graphical user interfaces.

The FFT benchmark from the SPLASH2 benchmark suite is listed under signal processing and communication. The FFT benchmark is an implementation of Fast Fourier Transform algorithm which is optimized to minimize interprocess communication\cite{scwSPLASH2Programs1995}. Signal processing and communication is one of the most common applications in an IoT network. FFT is an important Digital Signal Processing (DSP) algorithm which is required in communication of data over Software Defined Radios (SDR)\cite{taMicroprocessorArchIoT2015}.

These benchmarks, which require higher processing capabilities, are useful in determining the microarchitecture configurations of high-performance optimized host processor for the two-tiered heterogeneous processor architecture.

\subsubsection{Microarchitecture configurations for low-power optimized processors for IoT}
%
%

Table \ref{table_results_low_power} shows the microarchitecture configurations obtained for Cholesky and Radix benchmarks from the SPLASH2 benchmark suite. In these configurations, we note that for low-power optimized processor, the lowest operating frequency and core count are selected. This result can be interpreted intuitively, because high operating frequency and high number of cores in the processor increases the power consumption of the processor. We also note that these configurations have large L1-D cache sizes. This is because of the large workload offered by the test benchmarks. This is representative of the growing IoT ecosystem in which large volumes of data are gathered from a large number of sensing elements. The values of total power and execution times for microarchitecture configurations are also shown in Table \ref{table_results_low_power}. We observe that the power values are in the range of a hundred milliwatts and the execution time is in the range of a few hundred milliseconds. These values are within the operational requirements in most IoT deployments. These configurations implement the interface processors in the two-tiered heterogeneous processor architecture. With low-power requirements, these processors can always be operated in active mode, without impacting the power budget of IoT deployments

\subsubsection{Microarchitecture configuration for high-performance optimized processors for IoT}
\begin{table*}[!t]
\caption{Mircoarchitecture configuration for high-performance optimized processors for IoT}
\label{table_results_high_performance}
\centering
\begin{tabular}[width = \columnwidth]{| l | c | c | c | c | c |}\hline
\multirow{2}{*}{\textbf{Parameter Name}}          & \multicolumn{5}{ c |}{\textbf{Microarchitecture Configuration}} \\\cline{2-6}
                                 & \textbf{Blackscholes}   & \textbf{Freqmine} & \textbf{Facesim} & \textbf{Fluidanimate} & \textbf{FFT} \\\hline
Cores                            & 8 & 2 & 2 & 2 & 4 \\\hline
Frequency (MHz)                  & 3200 & 3200 & 3200 & 3200 & 3200 \\\hline
L1-I Cache Size (kB)         & 64 & 32 & 8 & 8 & 128\\\hline
L1-D Cache Size (kB)         & 128 & 128 & 64 &  64 & 32 \\\hline
L2 Cache Size (kB)           & 256 & 1024 & 1024 & 1024 & 512 \\\hline
L3 Cache Size (kB)           & 8192 & 2048 & 8192 & 4096 & 2048 \\\hline
\textbf{Total Power (W)}         & 4.549 & 1.565 & 1.546 & 1.546 & 2.563 \\\hline
\textbf{Execution Time (ms)}     & 28.1239 & 67.319 & 60.072 & 55.605 & 29.986 \\\hline
\end{tabular}
\vspace{-5mm}
\end{table*}
%
Table \ref{table_results_high_performance} shows the microarchitecture configurations obtained for Blackscholes, Freqmine, Facesim and Fluidanimate benchmarks from the PARSEC benchmark suite and the FFT benchmark from the SPLASH2 benchmark suite. We analyze the microarchitecture configurations obtained for these test benchmarks according to the categorization discussed in subsection \ref{results_application_IoT}. We observe that for data analysis and data mining applications, represented by the Blackscholes and Freqmine benchmarks, higher performance is achieved primarily by the increase in operating frequency. We note that the size of the L1-D cache for these applications is also high, which is because both are highly data-parallel benchmarks. The size of the L2 cache, for Blackscholes, and, L3 cache, for Freqmine, is also high which is also a result of data-parallelism in these benchmarks.
For graphics applications, represented by Facesim and Fluidanimate benchmarks, higher performance can again be attributed to increase in operating frequency. These benchmarks are also highly data-parallel which explains the large L1-D cache, L2 cache and L3 cache in the resulting microarchitecture configurations.
In signal processing and communication applications, represented by FFT benchmark, performance improvement, similar to other applications, is attained by increase in operating frequency. However, FFT requires a larger instruction cache as compared to larger data caches for other applications. Higher L1-I cache could be a result of the FFT benchmark being optimized for low interprocess communication.

The total power and execution time of each microarchitecture configuration is also listed in Table \ref{table_results_high_performance}. These configurations have high total power values in the range of one to a few watts but significantly low execution time values in the range of few tens of milliseconds. These configurations implement the host processor in the two-tiered heterogeneous processor architecture. Due to their high-power requirement, these processors are mostly kept in sleep mode and are activated intermittently for short durations to save energy and prolong battery life. Because these processors have shorter execution times, they can execute their tasks quickly and go to sleep thus, decreasing the duration that they are active.

%
%
\section{Conclusion and Future Work}
\label{conclusion_and_future_work}
In this paper, we proposed a temporally efficient design space exploration methodology for selecting microarchitecture configurations of processors for IoT. Our exploration methodology consisted of four phases. In the first phase, we determined best initial settings for tunable processor design parameters using initial one-shot search method. We also calculated the significance of each design parameter on the overall design in this phase. The results of this phase were used in the second phase to separate the processor design parameters into distinct search sets using an exploration threshold value supplied by the system designer. The third and the fourth phase of the methodology implemented exhaustive and greedy search methods to prune these search sets to determine the best microarchitecture configuration of the processor.

We tested our methodology over two design spaces, one for determining low-power optimized and the other for determining high-performance optimized processors for IoT. We validated the results obtained from our methodology by comparing with solutions obtained from fully exhaustive exploration of the design spaces. Our results revealed that our methodology obtained microarchitecture configurations close to within 2.23\%--3.69\% of the configurations obtained from fully exhaustive search. Our methodology only explored 3\%--5\% of the overall design space to determine these high quality solutions. This resulted in 24.16$\times$ average speedup on design space exploration as compared to the time required for fully exhaustive exploration.

We also described a two-tiered heterogeneous processor architecture for incorporating power-optimized performance in IoT objects. We used the results obtained from the evaluation of our design space exploration methodology to describe the two-tiered architecture. We categorized the test benchmarks into four different categories, relating them with possible IoT use cases and analyze microarchitecture configurations determined for these benchmarks to make our assertions on processors for IoT objects. We determined that for low-power optimization, microarchitecture configurations with lower core count and lower operating frequency are more suitable. For high-performance optimization, improvement in performance primarily results from increase in operating frequency. We also analyzed the cache hierarchy for different microarchitecture configurations and related them with the type and size of workloads offered by the test benchmarks.

In the future, we plan to investigate microarchitecture configurations of ultra-low power processors for IoT. We also intend to test our design space exploration methodology using standard IoT benchmarks. We also aim to improve our methodology by incorporating better optimization techniques like genetic and evolutionary algorithms and machine-learning. We also plan to study the practical applicability of the two-tiered heterogeneous processor model for processors for IoT objects, and, compare the model with processor architecture models currently in use in the IoT market.
%
%
{
\bibliographystyle{IEEEtran}
\bibliography{IEEEabrv,TETC_Final}
}

\begin{IEEEbiography}[{\includegraphics[width = 1in, bb = 0 -1 105 134]{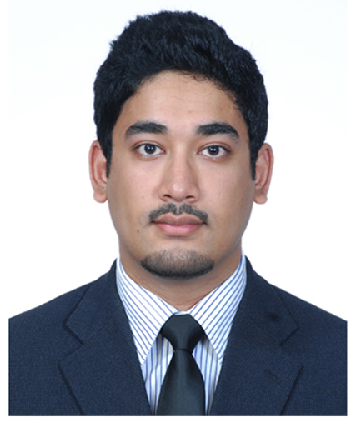}}]{Prasanna Kansakar} is a PhD student in the Department of Computer Science (CS) at Kansas State University (K-State), Manhattan, KS. His research interests include Internet of Things, embedded and cyber-physical systems, computer architecture, multicore, secure and trustworthy systems, and hardware-based security. Kansakar has an MS degree in computer science and engineering from the University of Nevada, Reno (UNR). He is a student member of the IEEE.
\end{IEEEbiography}

\begin{IEEEbiography}[{\includegraphics[width = 1in, bb = 0 -1 722 1010]{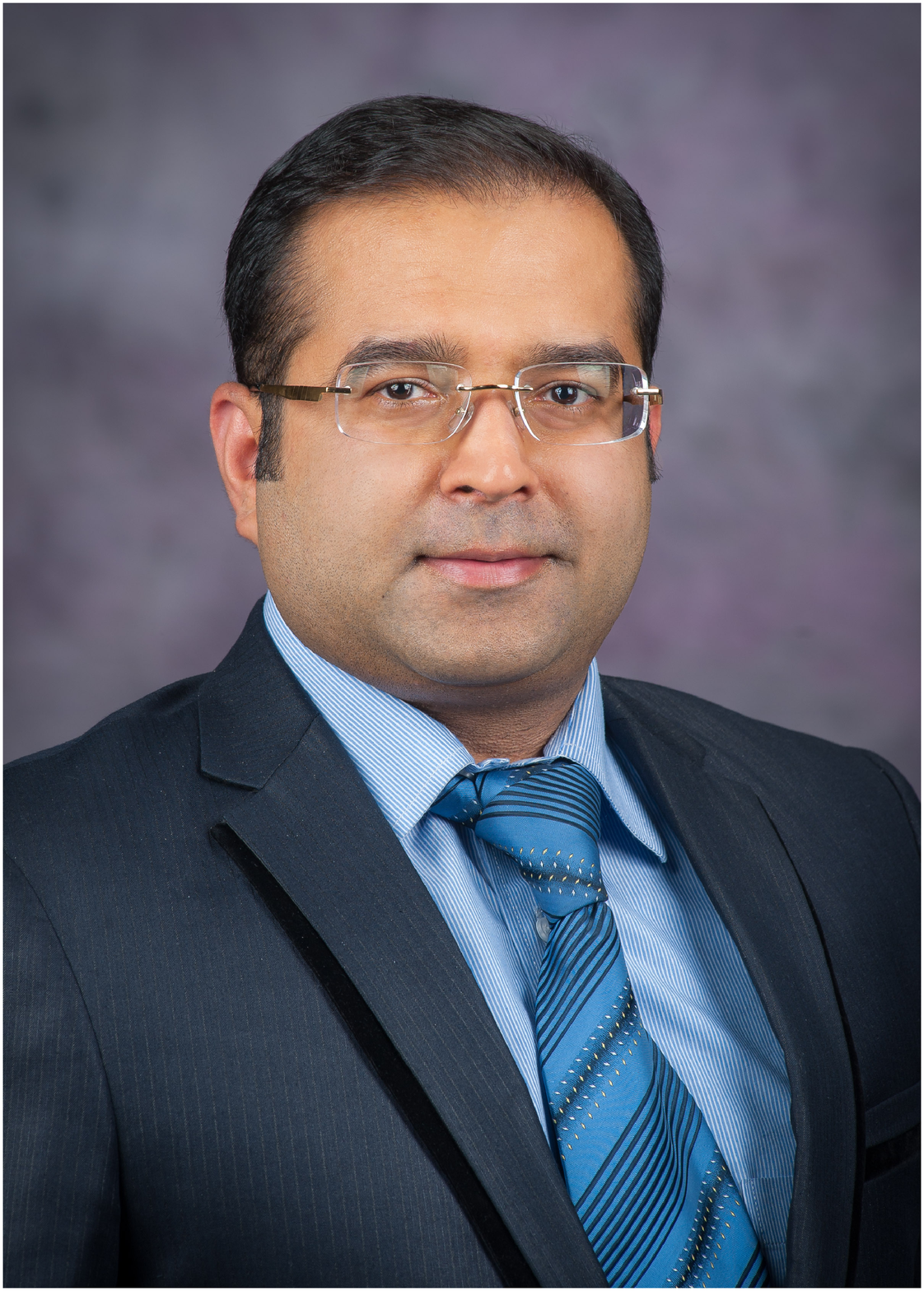}}]{Arslan Munir} is currently an Assistant Professor in the Department of Computer Science (CS) at Kansas State University (K-State). He holds a Michelle Munson-Serban Simu Keystone Research Faculty Scholarship from the College of Engineering. He was a postdoctoral research associate in the Electrical and Computer Engineering (ECE) department at Rice University, Houston, Texas, USA from May 2012 to June 2014. He received his M.A.Sc. in ECE from the University of British Columbia (UBC), Vancouver, Canada, in 2007 and his Ph.D. in ECE from the University of Florida (UF), Gainesville, Florida, USA, in 2012. From 2007 to 2008, he worked as a software development engineer at Mentor Graphics in the Embedded Systems Division.

Munir's current research interests include embedded and cyber-physical systems, secure and trustworthy systems, hardware-based security, computer architecture, multicore, parallel computing, distributed computing, reconfigurable computing, artificial intelligence (AI) safety and security, data analytics, and fault tolerance. Munir received many academic awards including the doctoral fellowship from Natural Sciences and Engineering Research Council (NSERC) of Canada. He earned gold medals for best performance in electrical engineering, gold medals and academic roll of honor for securing rank one in pre-engineering provincial examinations (out of approximately 300,000 candidates). He is a Senior Member of IEEE.
\end{IEEEbiography}
\fussy
}

\end{document}